\documentstyle[12pt,epsf,epsfig,psfig]{article}
\def\J{$J/\psi$}
\def\j{J/\psi}
\def\X{$\chi$}

\def\P{$\psi'$}

\def\U{$\Upsilon$}

\def\e{\epsilon}

\def\be{\begin{equation}}
\def\ee{\end{equation}}

\def\lsim{\raise0.3ex\hbox{$<$\kern-0.75em\raise-1.1ex\hbox{$\sim$}}}
\def\gsim{\raise0.3ex\hbox{$>$\kern-0.75em\raise-1.1ex\hbox{$\sim$}}}

\def\NP{{ Nucl.\ Phys.\ }}
\def\PL{{ Phys.\ Lett.\ }}
\def\PR{{ Phys.\ Rev.\ }}

\def\PRL{{ Phys.\ Rev.\ Lett.\ }}

\def\ZP{{ Z.\ Phys.\ }}
\def\EP{{ Eur.\ Phys.\ J.}}

\begin{document}

\noindent November 16, 2001 \hfill BI-TP 2001/30

\vskip 1.5 cm

\centerline{{\large{\bf Matter \& More in Nuclear
Collisions}}
\footnote{Concluding talk at
the {\it International Workshop on the Physics of the Quark-Gluon
Plasma}, \'Ecole Polytechnique, Palaiseau/France, Sept.\ 4 - 7, 2001}}

\vskip 0.5cm

\centerline{\bf Helmut Satz}

\bigskip

\centerline{Fakult\"at f\"ur Physik, Universit\"at Bielefeld}
\par
\centerline{D-33501 Bielefeld, Germany}

\vskip 0.5cm

\noindent

\centerline{\bf Abstract:}

\medskip

The aim of high energy nuclear collisions is to study the transition
from hadronic matter to a plasma of deconfined quarks and gluons. I
review the basic questions of this search and summarize recent
theoretical developments in the field.

\vskip 0.5cm

\noindent{\bf 1.\ New States of Matter}

\medskip

\begin{figure}[htb]
\centerline{\psfig{file=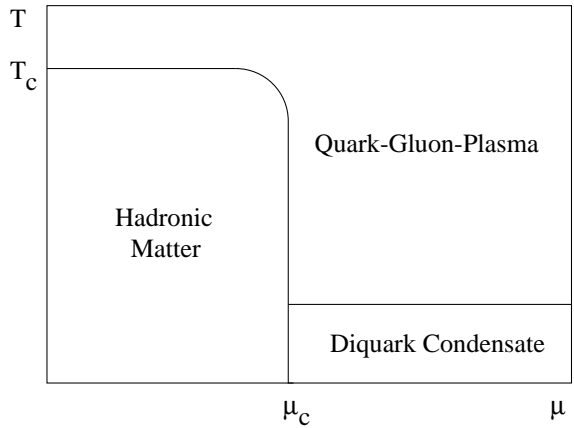,width=7.5cm}}
\caption{Phase diagram of strongly interacting matter.}
\label{phase}
\end{figure}

Statistical QCD predicts that high temperatures and baryon densities
will lead to new states of strongly interacting matter. Increasing $T$
at low baryon density transforms a meson gas into a deconfined plasma of
quarks and gluons (QGP); this transition has been studied extensively in
computer simulations of finite temperature lattice QCD. High baryon
densities at low $T$ are expected to produce a condensate of colored
diquarks. The resulting phase diagram in terms of temperature and 
baryochemical potential $\mu$ is schematically illustrated in
Fig.\ \ref{phase}, with hadronic matter as color insulator, the QGP as color
conductor, and the diquark condensate as color superconductor.

With high energy nuclear collisions, we want to study in the
laboratory the deconfinement transition and the properties of the QGP.
Hard probes, such as the production of quarkonia, open charm and
beauty, jets and photons are expected to provide information about the
hot early stages of the produced medium. The masses and decays of
different hadrons, their momenta, correlations and relative abundances
constitute the soft probes to study the later stages of the medium and
its freeze-out.

In this concluding talk, I will summarize some recent theoretical
developments in the field, without any claim to completeness. My
emphasis will be on concepts more than on specific models, and on
questions more than on answers.

\medskip

\noindent{\bf 2.\ Thermalization}

\medskip

Since the basic purpose of the experimental program is to produce
strongly interacting matter, it is of central importance to determine
if, how and when the non-thermal initial state of two colliding nuclei
becomes thermalized.

In a nuclear collision, the incoming nucleons or their secondaries can
interact with the other target nucleons. This results in nuclear
phenomena such as the Cronin effect, normal nuclear quarkonium
suppression or parton energy loss in normal nuclear matter. These
effects do not involve any new produced medium. To achieve that, the
secondaries coming from different sources must interact, a phenomenon
referred to as color interconnection, exogamous behavior or cross talk
\cite{exoga}.

A test for such cross-talk has been considered in $e^+e^-$ annihilation
into hadrons at $\sqrt s = 2M_W$, as studied in LEP experiments at CERN
\cite{LEP}. The reaction first leads to $W^+W^-$ production;
subsequently, one possibility is that each of the two $W$'s decays
into a $q \bar q$ pair, which then hadronizes. An
alternative channel has one of the two $W$'s undergo leptonic decay
into a neutrino and a lepton. In the case of cross talk, it is predicted
\cite{exoga} that for the resulting hadron multiplicities, $N_h(q_1\bar
q_1,q_2 \bar q_2) \not= 2N_h(q \bar q, \nu {\sl l})$; in addition, the
source radii obtained through HBT studies should be different in the
two channels. Neither of these predictions is supported by LEP data,
so that so far there is no evidence of cross talk shown by the hadrons
produced in high energy $e^+e^-$ annihilation. This also excludes
cross talk at earlier partonic stages.

\begin{figure}[htb]
\centerline{\psfig{file=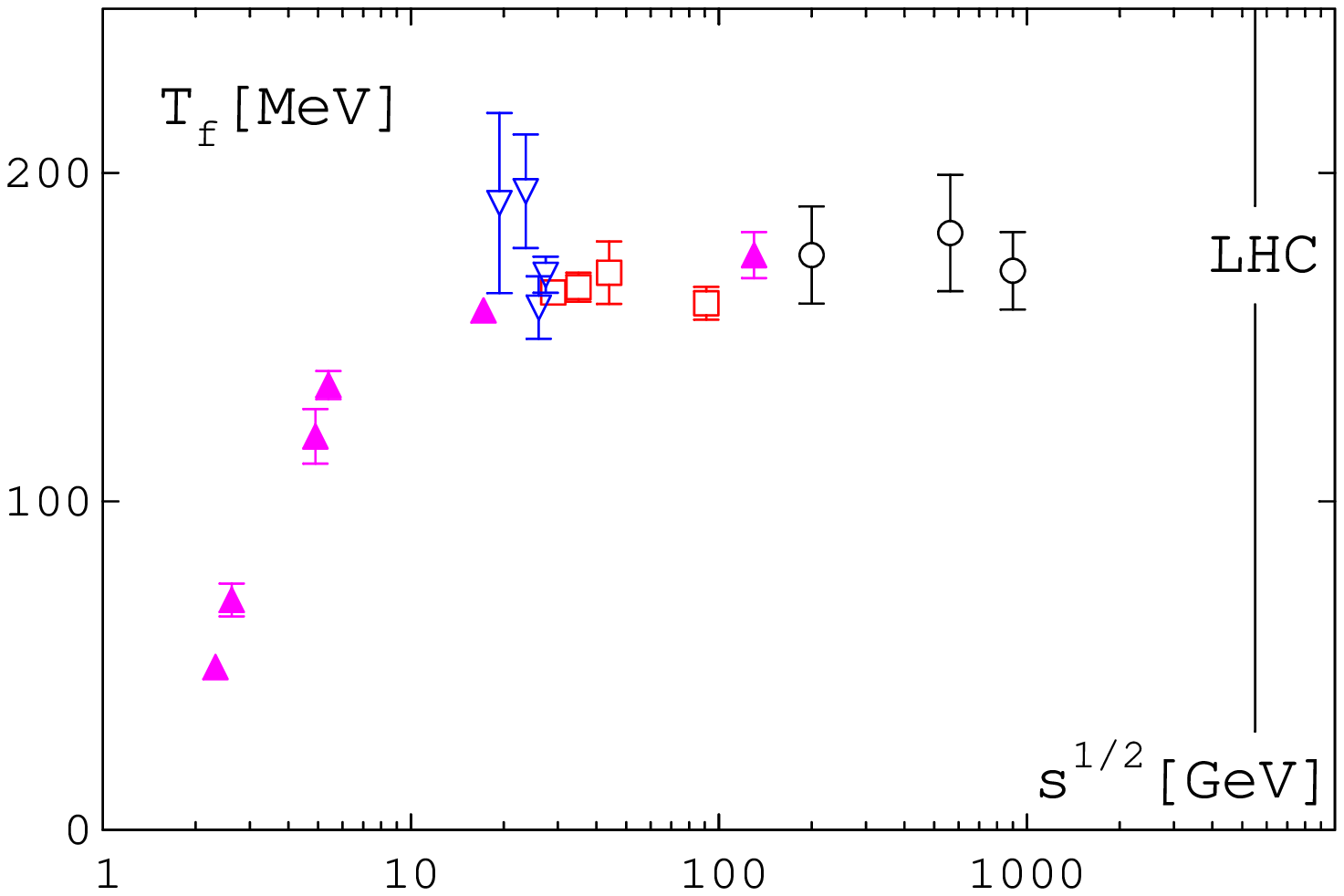,width=7cm,height=5cm}}
\vspace*{0.3cm}
\caption{Freeze-out temperatures for hadron resonances produced in 
nucleus-nucleus (filled triangles), $p\!-\!p$ (open triangles), 
$p\!-\!\bar p$ (circles) and $e^+e^-$ interactions (squares). Data are 
given in [3-5].}
\label{Tf}
\end{figure}

Do $AA$ collisions with their much higher density of superimposed
interactions lead to cross talk? We address this question by
looking at hadron abundances. It is found that these are quite well
described by the predictions of an ideal resonance gas \cite{RG},
parametrized in grand-canonical form by a freeze-out temperature $T_f$
and a baryochemical potential $\mu_B$. With increasing collision
energy $T_f$ converges to about 170 MeV (Fig.\ \ref{Tf}). This alone
does not, however, allow us to conclude that we have indeed obtained a
thermal system with full cross-talk. It is known \cite{beca} that also
the elementary hadroproduction processes initiated by $e^+e^-$
annihilation or $pp/ p{\bar p}$ scattering lead to thermal hadron
abundances, with freeze-out temperatures which agree very well with
those observed in $AA$ collisions; they are included in Fig.\ \ref{Tf}.
There is, however, one important difference: in the elementary
reactions, there are fewer strange hadrons than predicted by a
grand-canonical resonance gas, while $AA$ collisions do not show such a
strangeness reduction. The change in relative strangeness production,
referred to as strangeness suppression or enhancement, depending on the
point of view, has found a very natural explanation in the observation
\cite{redlich} that in elementary processes, with rarely more than one
$s \bar s$ pair per interaction region, strangeness conservation has to
be taken into account exactly and not just ``on the average", as
implied in a grand-canonical formulation. It had in fact been observed
long ago that the exact local conservation of quantum numbers can
reduce the relative production rates by several orders of magnitude in
comparison to the average grand-canonical rates \cite{hage}.

The transition from exact to average (grand-canonical) strangeness
conservation does imply, however, that the $AA$ collisions behave as
one large system, not as a sum of many elementary collisions. The
strange hadrons produced in one elementary collision of an $AA$
interaction must be aware of the strange hadrons produced in other $NN$
collisions, if a grand-canonical description is valid. So at least at
the hadronization stage of the medium produced in nuclear collisions,
there is cross talk -- it makes sense to speak of a large-scale
hadronic medium.

One remaining question in this context is the correct treatment of
hidden strangeness; if the $s \bar s$ nature of the $\phi$ is not taken
into account, its production rates are quite generally overpredicted.

In addition, there remains the tantalizing question of why the hadron
abundances in elementary processes like $e^+e^-$ or $pp/ p{\bar p}$
already follow the pattern predicted by an ideal resonance gas. It might
just indicate that the non-perturbative hadronization of quarks and
gluons simply proceeds such as to maximize the entropy: partons
hadronize into the states of a resonance gas with largest phase space.
The associated temperature is then the limiting temperature of such a
system, the Hagedorn temperature, and hence is universal.

Since the initial structure of e.g.\ $e^+e^- \to$ hadrons is that of a
fast $q \bar q$ pair emitting gluons which later hadronize, the
annihilation process is certainly not thermal in its early stages.
Thermal hadron abundances thus do not imply that the previous partonic
system was already thermal.

Partonic cross talk is the basic input of most parton cascade models
and thus has been considered in the corresponding codes for quite some
time. Its role in the establishment of parton thermalization has
recently been addressed in an interesting conceptual treatment
\cite{BMSS}. The two
incoming nuclei can be viewed in the central rapidity region (where the
valence quarks are unimportant) as gluon beams in which each gluon has a
transverse radius $r_g \sim 1/k_T$ determined by its transverse momentum
$k_T$. The geometric interaction cross section for gluon of sufficient
hardness, $k_T \geq q_0$, is $\sigma_g \sim \alpha_s(q_0) r_g^2$. We
would like to know when these gluons of sufficient hardness begin to
overlap in the transverse area determined by the nuclear radius $R_A$.
Full cross talk evidently occurs when a gluon on one side of the
nuclear disk $\pi R_A^2$ is connected by overlapping gluons to a partner
on the other side of the disk. This starts at the percolation point
\be
N_g \sigma_g \simeq \pi R_A^2,
\label{2.1}
\ee
where $N_g(k_T) = A xg(x)$ denotes the number of gluons as determined
from the gluon distribution function $g(x)$ at central Bjorken $x \sim
\sqrt s$. From Eq.\ (\ref{2.1}) we thus obtain
\be
q_s^2 \sim \alpha_s A^{1/3} xg(x)
\label{2.2}
\ee
for the saturation momentum $q_s$. When $k_T >> q_s$, the gluons form a
dilute and hence disjoint system in the transverse plane, without
cross talk. At $k_T=q_s$ percolation sets in and the gluons form a
connected interacting system, with full cross talk. For such a system,
one can estimate that thermalization occurs after a time
\be
\tau_0 \sim \alpha_s^{-13/5} q_s^{-1},
\label{2.3}
\ee
leading to a thermal gluon medium of temperature
\be
T_0 \sim \alpha_s^{2/5} q_s.
\label{2.4}
\ee
From deep inelastic scattering one has $xg(x) \sim (\sqrt s)^{\lambda}$,
with $\lambda \simeq 0.2$. Together with Eq.\ (\ref{2.2}), this implies
that large nuclei (large A) and/or high collision energies (large $\sqrt
s$) lead to early parton thermalization and a hot QGP. A very important
task for theory is clearly to turn these conceptual considerations into
a quantitative formalism.

\medskip

\noindent{\bf 3.\ Hadrons in Matter}

\medskip

Do $AA$ collisions produce an interacting hadronic medium, or does the
earlier partonic state hadronize directly into an ideal resonance gas?
That is the main question to be addressed here. It is of particular
interest in view of chiral symmetry restoration. For temperatures $T
< T_c$, the massless quarks of the QCD Lagrangian ${\cal L}_{\rm QCD}$
``dress" themselves through gluon interactions to become constituent
quarks with an effective mass $M_q \sim$ 0.3 - 0.4 GeV, thereby
spontaneously breaking the chiral symmetry of ${\cal L}_{\rm QCD}$. For
$T=T_c$,
chiral symmetry is restored and $M_q \to 0$; we have here assumed a
system of vanishing baryon number density, for which deconfinement and
chiral symmetry restoration coincide at $T=T_c$. Since the vector meson
mass $M_{\rho} \simeq 2 M_q$, the behavior of $M_{\rho}(T)$ in an
interacting hadronic medium for $T \to T_c$ would be a way to study the
onset of chiral symmetry restoration.

The in-medium behavior of hadron masses can be calculated in finite
temperature lattice QCD. First studies addressed the temperature
dependence of the screening mass in quenched QCD. Below $T_c$, they showed 
very little $T$-dependence; but in view of the noted simplifications, they
are presumably not really conclusive. Today it is possible to calculate
the actual pole mass, but with present computer performance still only
for the quenched case \cite{Wetzorke}.
The advent of more powerful computers in the next
2 - 3 years should, however, lead to such calculations in full QCD with
light quarks. The present quenched studies show that above $T_c$, there
are no more mesonic bound states; scalar and pseudoscalar correlations
agree, indicating chiral symmetry restoration. Below $T_c$, the results 
for the pole masses also show rather little $T$-dependence; however, this
may well be an artifact of quenching, as the following considerations
seem to indicate.

\begin{figure}[htb]
\centerline{\psfig{file=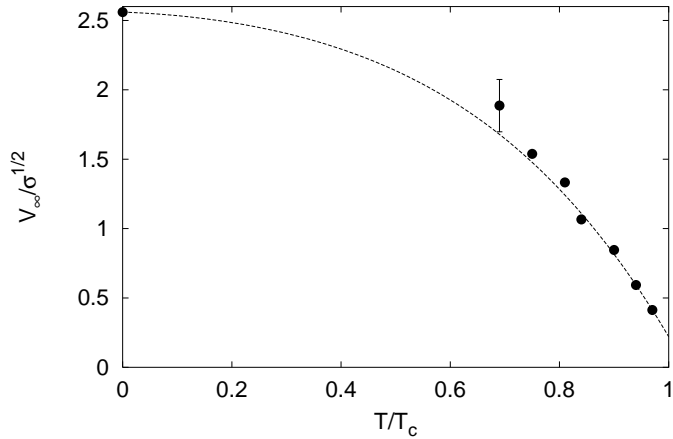,width=9cm}}
\vspace*{-0.1cm}
\caption{Temperature dependence of the string breaking potential 
$V_Q(T,\infty)$ in units of the string tension $\sigma$.} 
\label{plateau}
\end{figure}

Lattice studies of the heavy quark potential $V_Q(T,r)$ in
full QCD (with light dynamical quarks) show at all temperatures $T$
string breaking in the large distance limit $r \to \infty$. At $T=0$,
the string connecting two heavy color charges should break when its
energy surpasses the mass of a typical light hadron, or equivalently
when
\be
V_Q(T,r=\infty) \simeq 2 M_q(T).
\label{3.1}
\ee
Hence $M_q(T)$ can be determined in finite temperature lattice
calculations of $V_Q(T,r)$. Note, however, that the constituent quark 
mass here is obtained from a heavy-light meson and could contain some
dressing of the heavy quark; hence it need not coincide fully
with that from a light-light meson such as the $\rho$. Heavy quark
potential studies have recently been
carried out for $N_f = 2$ and 3 for a range of different quark masses
\cite{Peikert}. They show in particular that
\begin{itemize}
\vspace*{-0.1cm}
\item{for $m_q~ \lsim~ 0.4~T \simeq 60$ MeV, the dependence of $V_Q$
on $m_q$ becomes negligible, indicating that the chiral limit is
reached;}
\vspace*{-0.1cm}
\item{string breaking occurs earlier (at smaller $r$) with increasing
temperature (Fig.\ \ref{3.1});}
\vspace{-0.1cm}
\item{from $V_Q(T=0, r=\infty) \simeq 1$ GeV it follows that
$M_q(T=0) \simeq 0.5$ GeV, while for $T \to T_c$, $V_Q(T)$ and
hence also $M_q(T)$ vanish.}
\vspace*{-0.1cm}
\end{itemize}
The resulting temperature dependence of the heavy quark potential is
shown in Fig.\ \ref{plateau}. It is seen that the approach of chiral
symmetry restoration leads to a pronounced variation of $V_Q(T,\infty$)
and hence of $M_q(T)$ with
$T$. This suggests that also the mass of the $\rho$-meson should show
such a temperature variation, in contrast to the present pole-mass
results from quenched lattice QCD. A study of the temperature
dependence of hadron masses in unquenched lattice calculations would
thus be of great interest.

\medskip

\noindent{\bf 4.\ Partons in Matter}

\medskip

Since the energy loss of a fast parton passing through a medium will
depend on the nature of this medium, jet quenching should provide a tool
to specify the state of matter produced in nuclear collisions. At lower
momenta, the energy loss can occur through ionisation of the
constituents of the medium; at high momenta, gluon radiation of the
passing parton is the main mechanism.

\begin{figure}[htb]
\mbox{
\hskip0.5cm\epsfig{file=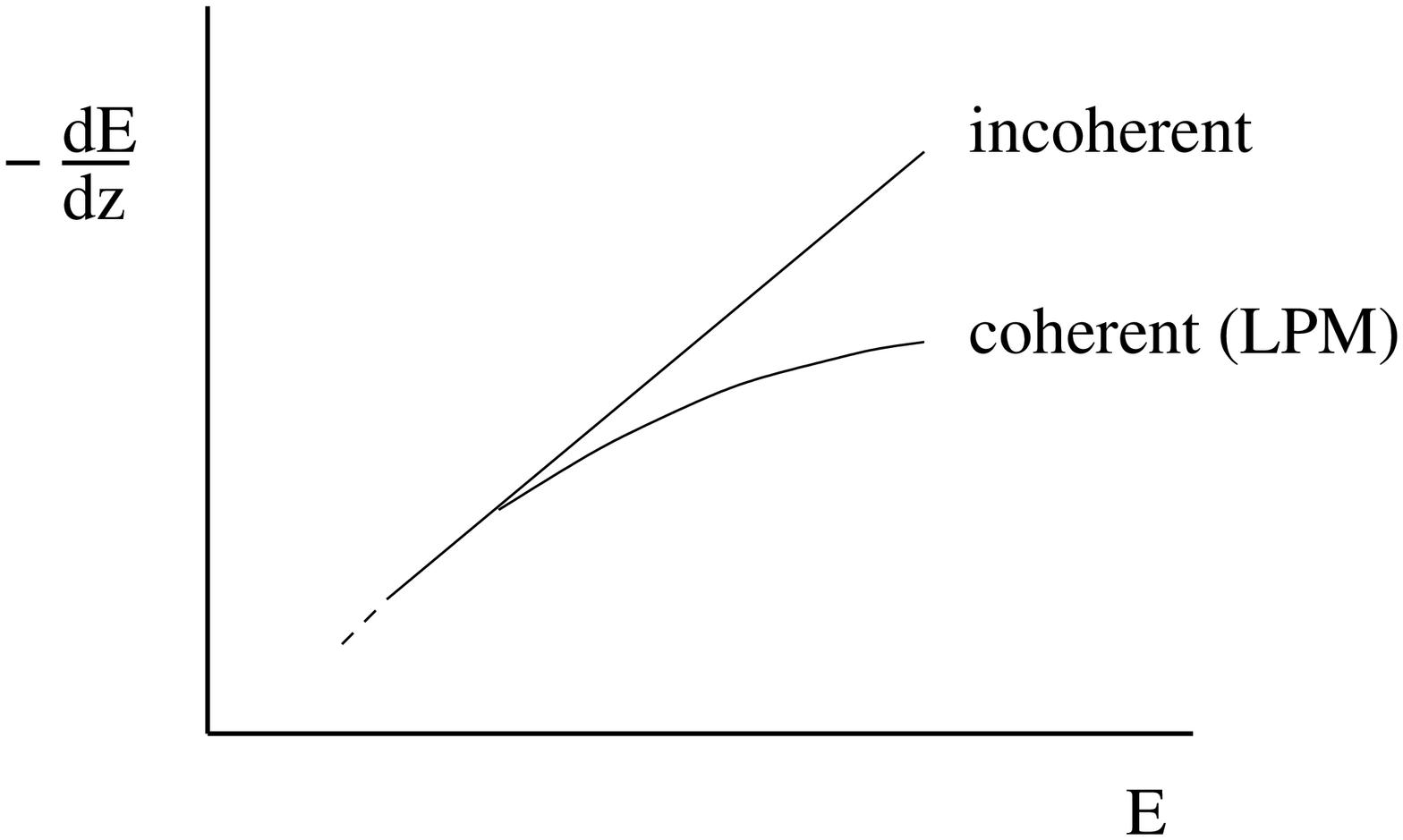,width=5cm,height=6cm,angle=-90}
\hskip2cm
\epsfig{file=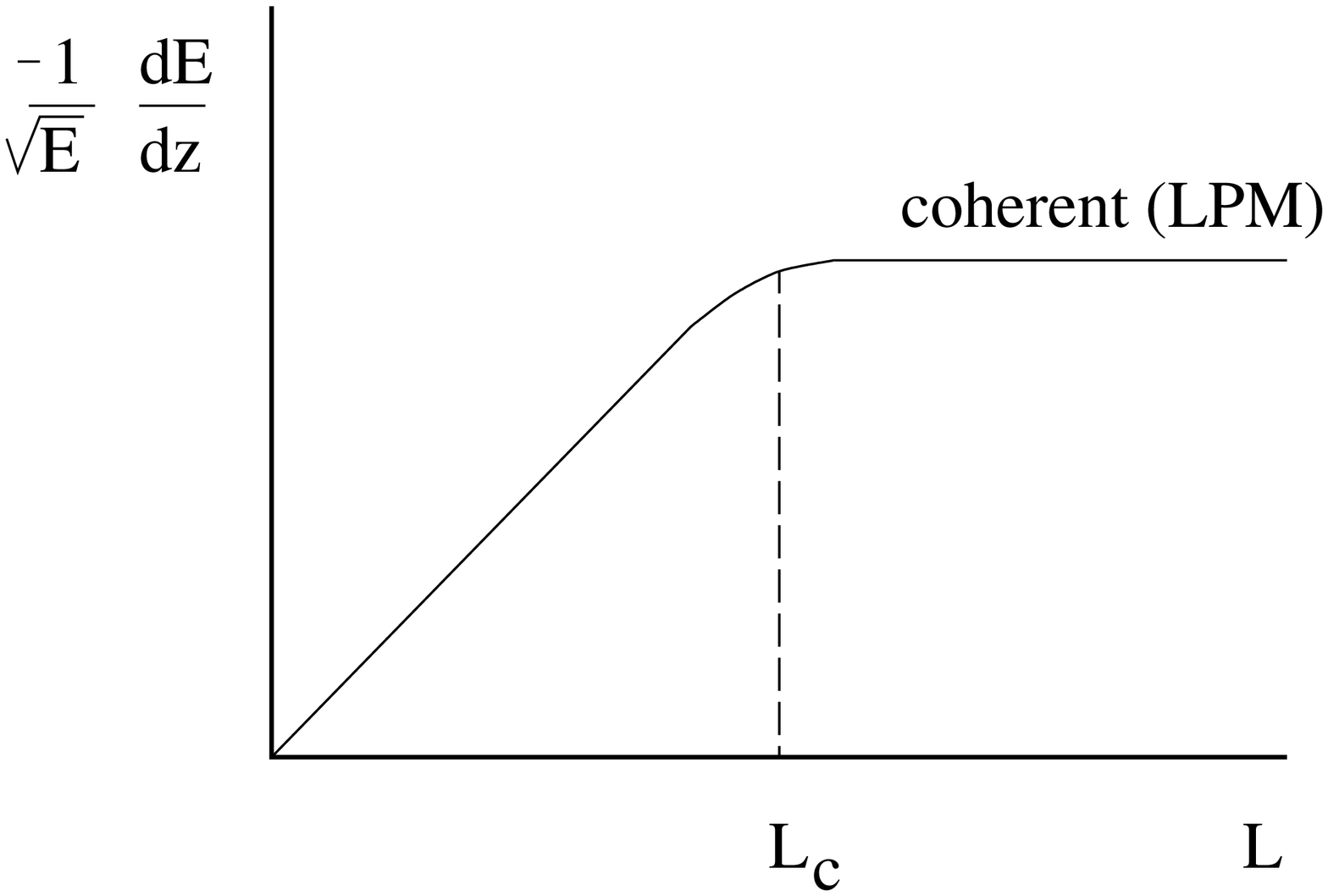,width=5cm, height=6cm,angle=-90}}
\vskip0.3cm
\caption{Parton energy loss through coherent vs. incoherent scattering 
(left) and through coherent scattering in a medium of finite size (right). }
\label{parton}
\end{figure}

The crucial feature in radiative energy loss is the formation time
$t(k)$ or the formation length $z(k)$ for a gluon of momentum $k$,
compared to the intrinsic scales of the medium: the mean free path
$\lambda$, the mean distance $d$ between scattering centers and the
overall linear size $L$ of the medium. For $\lambda > d > z(k)$,
the radiated gluons see independent charges and the scattering is
incoherent. For $d < \lambda < z(k)$, there is coherent scattering of
the nascent gluon with several scatterers, leading to destructive
interference; this is the so-called Landau-Pomeranchuk-Migdal (LPM)
effect which reduces the energy loss. In Fig.\ \ref{parton} we compare
schematically the incoherent form $dE/dz \sim - E$, where $E$ is the
parton energy, to the LPM form $dE/dz \sim - \sqrt E$. In a medium of
small linear size, $L < L_c$, there is a further finite size reduction,
leading to \cite{BDMPS}
\be
-{dE \over dz} \simeq {3\alpha_s \over \pi}
\cases{(\mu^2 E /\lambda)^{1/2}&, $L > L_c$ \cr
(\mu^2 / \lambda)~L &, $L < L_c$}
\label{4.1}
\ee
for the energy loss in a quark-gluon plasma. Here
$L_c \equiv (E \lambda/\mu^2)^{1/2}$ denotes the limiting length scale
and $\mu^{-1}$ the screening length of the medium.

For a parton traversing a QGP at temperature $T=250$ MeV over a length
of $L=10$ fm, Eq.\ (\ref{4.1}) leads to an energy loss of about 30 GeV;
this is to be compared to indications that cold nuclear matter
of the same size would only result in an energy loss of 2 GeV \cite{schiff}.
In contrast to the QGP results, the value for a normal nuclear medium
is really only an estimate, and thus only jet production data from $pA$
collisions can provide a reliable basis for comparison.

Another problem is immediately evident from the QGP calculation. What
transverse momenta are really needed to specify a jet? Present RHIC data
stop around at leading particles of some 5 GeV, which presumably is well
below the value for the jets assumed in the QCD studies.

\begin{figure}[htb]
\centerline{\psfig{file=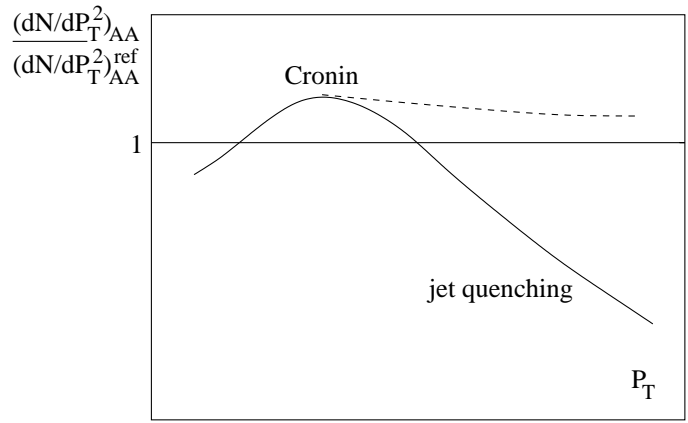,width=9cm}}
\vspace*{0.6cm}
\caption{Expected transverse momentum distribution from $AA$ collisions 
normalized
to a reference distribution interpolating from a wounded nucleon to a 
binary collision model (see Eq.\ \ref{4.2}).}
\label{p-T}
\end{figure}

We add a comment here on the interpretation of the RHIC results on high
$p_T$ hadrons. The measured spectra have to be compared to some
reference spectrum in order to look for a possible quenching, and this
is generally based on binary collisions: the spectra from central $AA$
collisions are normalized to $pp$ (or peripheral $AA$) data multiplied
by the number of binary collisions. The resulting ratio (see Fig.\
\ref{3.1}) is well below unity in the entire range $0 ~\lsim~ p_T~
\lsim~ 5$ GeV. If one would instead use the number of wounded nucleons
as reference, the corresponding ratio will be larger than unity for
almost all values of $p_T$. From multiplicity studies it is clear that
the overall data, dominated by low to intermediate $p_T$, are well below
what is expected from binary collision scaling. Hence in order to
obtain a reliable reference, one should use an interpolating form of
the type
\be
\left({dN \over dp_T^2}\right)_{AA}^{\rm ref} = \left[
N_w \left(1 -  {p_T^2 \over a + p_T^2} \right) +
N_c \left( {p_T^2 \over a + p_T^2} \right) \right]
\left({dN \over dp_T^2}\right)_{pp},
\label{4.2}
\ee
where $N_w$ denotes the number of wounded nucleons and $N_c$ the number
of binary collisions; the parameter $a$ determines the relative
importance of the two types of production mechanisms. Instead of being
set to zero, as in present studies, it should be choses such as to
correctly reproduce the measured multiplicity. This is expected to lead
to a behavior like that shown in Fig.\ \ref{p-T}, with a Cronin-like
pattern at relatively low $p_T$ followed by a suppression below unity,
and one could then clearly define quenching effects.

The last point to be addressed in this section concerns the radiative
energy loss of heavy quarks traversing a QGP. It was noted \cite{DK}
that for massive quarks the gluon emission suffers a `dead-cone'
effect, which suppresses radiation for forward angles
\be
\theta ~\lsim ~M_Q/\sqrt{P_Q^2 + M_Q^2},
\label{4.3}
\ee
where $M_Q$ denotes the mass of the heavy quarks. This radiation
suppresion in turn reduces the energy loss of heavy quarks and thus
predicts an increase of the ratio $D/\pi$ for high $p_T$.

\medskip

\noindent{\bf 5. Quarkonia in Matter}

\medskip

The essential feature that distinguishes the quarkonium ground states
\J~and \U~from the normal light hadrons is their much smaller radius
(about 0.2 fm for the \J~and about 0.1 fm for the \U), due to the much
higher bare quark mass ($m_c \simeq 1.4$ GeV, $m_b \simeq 4.5$ GeV).
Their binding is thus largely due to the Coulombic part of the QCD
potential $\sigma r - \alpha/r$; the string tension $\sigma$ does not
matter very much. Equivalently, the gluon dressing which makes massive
constituent quarks out of the almost massless light quarks (with $M_q
\simeq 0.3 - 0.4$ GeV) has little effect on the heavy quarks. In a
medium approaching the deconfinement point, for $T \to T_c$, the string
tension vanishes: $\sigma(T) \to 0$, as does the constituent quark mass
because of chiral symmetry restoration, $M_q \to 0$. As a consequence,
light and light-heavy hadrons disappear, but sufficiently tightly bound
quarkonia will persist even above $T_c$ and can thus serve as probes of
the quark-gluon plasma. These arguments also suggest that quarkonium
masses decrease less with temperature than the masses of the open
charm or beauty mesons $D$ and $B$. As a consequence, the open charm
threshold can in a hot medium fall below the mass of previously stable
higher excited charmonium states and thus allow their strong decay, and
similarly for bottomonia.

We therefore want to compare $2M_D(T)$ with $M_i(T)$, where $i$
specifies \P, \X~ and \J, as well as the corresponding $b$-quark states.
For the masses of the open charm/beauty states, we make use of the
lattice studies already introduced in section 3. The string breaking
potential introduced there determines with $V(T,r=\infty) \simeq
2 M_q(T) \simeq 2(M_D - m_c)$ effectively the $D$-mass. The quarkonium
masses can be obtained by solving the Schr\"odinger equation with the
potential $V(r,T)$ determined in the same lattice studies. Comparing the
temperature dependence of the light-heavy masses to that of the
quarkonium states shows two distinct types of behavior \cite{DPS1,DPS2}.

\begin{figure}[htb]
\mbox{
\epsfig{file=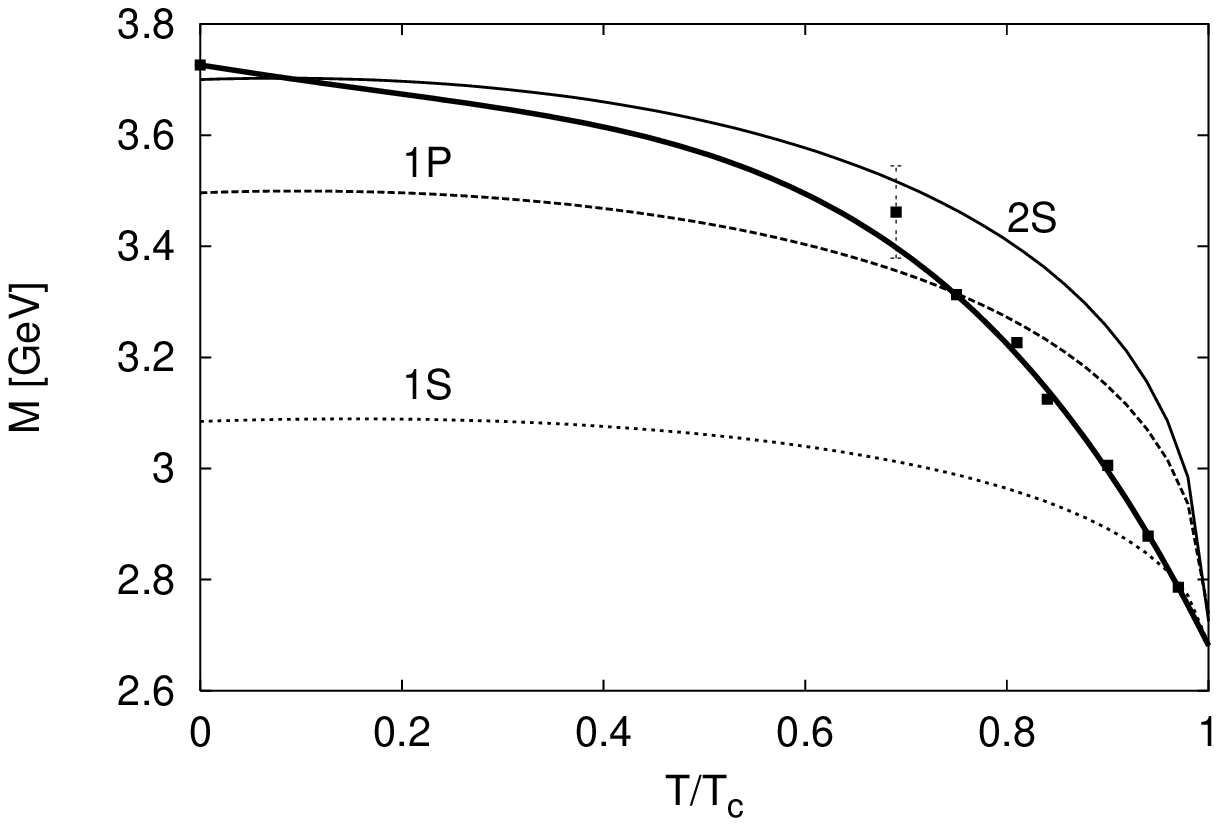,width=7cm,height=6cm}
\hskip0.9cm
\epsfig{file=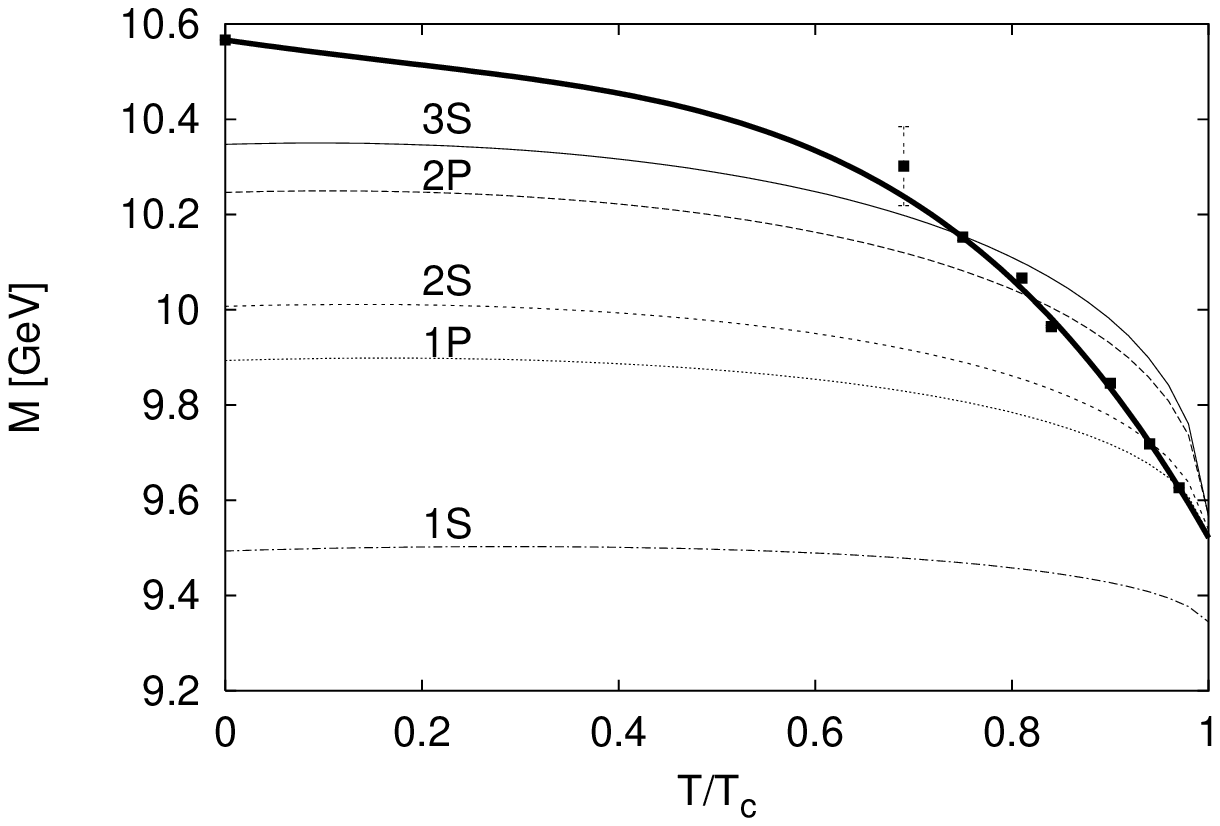,width=7cm, height=6cm}}
\vskip0.5cm
\vspace*{-0.5cm}
\caption{Temperature dependence of (left) open charm and (right) open beauty
masses (heavy lines) vs.\ charmonium and bottomonium masses.}
\label{mass}
\end{figure}

In Fig.\ \ref{mass} we see that with increasing temperature $2M_D$
and $2M_B$ indeed drop below the masses of the highest excited states,
\P~and \X$_c$ for charmonia, \U'' and \X$_b'$ for bottomonia,
respectively, before the deconfinement point is reached. These states
thus disappear in a hot hadronic medium through in-medium decay into
open charm/beauty. If the dissociation thresholds are experimentally
measured, they thus specify the temperature of the hot but still
confined system at four different points, tracing out the approach of
chiral symmetry restoration.

\begin{figure}[htb]
\vskip 0.5cm
\centerline{\psfig{file=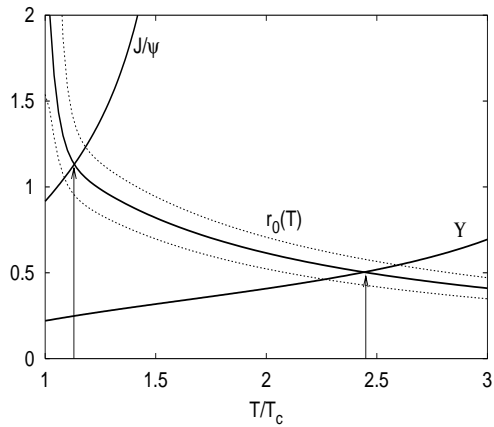,width=7cm,height=6cm}}
\vspace*{-0.2cm}
\caption{Temperature dependence of the screening radius vs.\ the 
bound state radii of \U~ and \J, in fm.}
\label{screen}
\end{figure}

The mass gaps of \J, \U', \X$_b$ and \U~at $T=0$ are much larger than
$\Lambda_{\rm QCD}$, so that we expect them to survive deconfinement and
be dissociated only by color screening in the quark-gluon plasma, as
originally proposed for the \J~\cite{Matsui}. This dissociation sets in
when the intrinsic scale of the quarkonium, its radius, falls below the
screening radius as the scale characterizing the medium. In Fig.\
\ref{screen} this effect is seen to occur for the \U~ at $T \simeq 2.3~
T_c$. The other mentioned states persist up to about $T_c$ when
compared to open charm/beauty masses (see Fig.\ \ref{mass}); in a
screening approach, they are dissociated in a QGP just slightly above
$T_c$ (see Fig.\ \ref{screen} for the \J). Given the accuracy of the
present lattice results near deconfinement, and in view of the possible
break-down of a Schr\"odinger equation near $T_c$, we can thus only
conclude that \J, \X$_b$ and \U' are dissociated approximately at $T_c$.

We thus obtain a thermal quarkonium dissociation pattern which indeed is
very similar to that provided by the spectral lines from stellar matter
\cite{Kajantie}. The suppression thresholds for \P,  \X$_c$, \U'' and
\X$_b'$ specify a hot hadronic medium different temperatures; when the
\J, \U' and \X$_b$ disappear, deconfinement is reached, and the
dissociation point of the \U~indicates a hot QGP, with $T>2~ T_c$.

The observed production of \J~and \U~occurs in part through the decay of
higher excited states, such as \X$_c \to \j$; the respective fractions
of the different contributions are known experimentally or can be
determined from data \cite{DPS2}. Since such decays take place far
outside the interaction region, the produced medium sees and suppresses
the different `parent' states. This leads to the well-known sequential
suppression pattern \cite{GS2,DPS2} which distinguishes thermal
threshold behavior such as deconfinement form dissociation by hadronic
comover scattering (see \cite{Vogt} for a survey). Perhaps the most
interesting feature which has so far emerged from nuclear collision
studies is the observation of just such a multistep structure of
\J~suppression \cite{NA50}. Future experiments, both at CERN (NA60) and
at RHIC, will undoubtedly provide further details to check if this
structure is indeed due to sequential quarkonium suppression.

Our considerations so far have ignored possible fluctuations of the
medium. To illstrate, we note that for $T>0$ the mass of the $D$ and to
a lesser extent the mass of the \X$_c$ will fluctuate around the values
we have here calculated. Instead of a $\delta$-function, we will have a
peak with a certain width (collision broadening), which in principle
can be provided by lattice calculations. For a conclusive study of
sequential quarkonium suppression in nuclear collision this would
seem a prerequisite.

A second open question concerning applications to experiment is a
reliable determination of the energy densities or temperatures attained
there. All present lattice studies find $T_c \simeq 0.15 - 0.20$ GeV for
the deconfinement temperature; the corresponding energy density is
$\e(T_c) \simeq 1$ GeV/fm$^3$, although it then grows quickly to values
near the Stefan-Boltzmann limit, so that $\e(1.1~T_c) \simeq 2$
GeV/fm$^3$. Presently quoted values for the energy densities in $Pb-Pb$
collisions at the CERN-SPS are in the range 2 - 3.5 GeV/fm$^3$; they
are based on Bjorken's estimate, which for central collisions gives
\be
\e \simeq \left({dN_h \over dy} \right)_{y=0} {p_0 \over \pi R_A^2
\tau_0},
\label{5.1}
\ee
where $dN_h/dy$ denotes the multiplicity and $p_0$ the average energy of
the produced hadrons, $R_A$ the nuclear radius and $\tau_0 \simeq 1$ fm
some average formation time of the medium. Obviously the choice of
$\tau_0$ is rather crucial, and a cross check of the reliability of the
resulting estimates would thus seem very necessary.

\bigskip

\noindent {\bf Summary}

\medskip

\begin{itemize}

\vspace*{-0.3cm}

\item{Hadron abundances, in nuclear collisions as well as in elementary
interactions, follow the pattern of an ideal resonance gas. The
strangeness suppression observed in elementary processes appears
accountable through exact strangeness conservation. The observed energy
independent freeze-out temperature $T_f \simeq 170$ MeV seems to
reflect critical features.}
\vspace*{-0.3cm}
\item{Finite $T$ lattice studies of the heavy quark potential $V(T,r)$
show a significant variation of the string breaking energy $V(T,\infty)$
for $T \to T_c$. This could be an indication for a similar temperature
variation of light hadron masses in full QCD.}
\vspace*{-0.3cm}
\item{Fast partons passing through a QGP suffer a considerable energy
loss, which should be observable for sufficiently hard jets or their
decay products. Present RHIC data require a reference distribution
interpolating from a wounded nucleon to a binary collision form.}
\vspace*{-0.3cm}
\item{Quarkonia in hot matter can be dissociated by two distinct
mechanisms. Higher excited states decay strongly into open charm/beauty
mesons when the masses of the latter decrease as the system approaches
chiral symmetry restoration. More tightly bound lower states survive
up to deconfinement and are subsequently dissociated by color screening
in the hot QGP.}

\end{itemize}

\medskip

\noindent{\bf Acknowledgements}

\medskip

It is a pleasure to thank many colleagues for helpful comments and 
suggestions; particular thanks go to R. Baier, S.\ Digal, F.\ Karsch, 
D. Kharzeev, P.\ Petreczky and K.\ Redlich. The financial support
of the German Ministry of Science (contract 06BI902) and of the GSI
(contract BI-SAT) is gratefully acknowledged.

\end{document}